\documentclass{svjour8}                    
\smartqed  
  \usepackage{natbib}
  \usepackage{rotating}
\usepackage{caption}
  \usepackage{floatpag}
  \rotfloatpagestyle{empty}
  \usepackage{wrapfig}
\usepackage{sidecap}
\usepackage{enumerate}
\usepackage{feyn,feynmp}
\usepackage{dsfont}

\usepackage{graphicx,tikz}

\newcommand{\be}{\nopagebreak[3]\begin{equation}}
\newcommand{\ee}{\end{equation}}
\newcommand{\ba}{\nopagebreak[3]\begin{eqnarray}}
\newcommand{\ea}{\end{eqnarray}}

\newcommand{\bc}{}

\begin{document}
\title{Relational Quantum Cosmology}

\subtitle{Contribution to the forthcoming book on Philosophy of Cosmology edited by K. Chamcham, J. Barrow, J. Silk and S. Saunders for Cambridge University Press.
}
\titlerunning{Discreteness and Relationalism} 

\author{Francesca Vidotto}
\institute{
Radboud University, Institute for Mathematics, Astrophysics and Particle Physics, 
Mailbox 79, P.O. Box 9010, 6500 GL Nijmegen, The Netherlands\\\email{fvidotto@science.ru.nl}
}

\date{August 3, 2015}

\maketitle

\begin{abstract}
The application of quantum theory to cosmology raises a number of conceptual questions, such as the role of the quantum-mechanical notion of ``observer'' or the absence of a time variable in the Wheeler-DeWitt equation.  I point out that a relational formulation of quantum mechanics, and more in general the observation that evolution is always relational, provides a coherent solution to this tangle of problems. 
\end{abstract}

\section{Conceptual problems in quantum cosmology} 

A number of confusing issues appear when we try to apply quantum mechanics to cosmology.   Quantum mechanics, for instance, is generally formulated in terms of an observer making measurements on a system. In a laboratory experiment, it is easy to identify the system and the observer: but what is the observer in quantum cosmology? Is it part of the same universe described by the cosmological theory, or should we think at it as external to the universe? Furthermore, the basic quantum dynamical equation describing a system including gravity is not the Schr\"{o}dinger equation, but rather the Wheeler-DeWitt equation, which has no time parameter: is this related to the absence of an observer external to the universe? How do we describe the quantum dynamics of the universe without a time variable in the dynamical equation and without an observer external to the universe?

I suggest that clarity on these issues can be obtained by simply recognising the relational nature of quantum mechanics and more in general the relational nature of physical evolution.  In the context of quantum theory, this nature is emphasised by the so called \emph{relational} interpretation of quantum mechanics [\cite{Rovelli:1995fv}].  The relational nature of evolution, on the other hand, has been pointed out by the partial-observable formulation of general covariant dynamics [\cite{Rovelli:2001bz}].  

A central observation 
is that there is a common confusion between two different meanings of the expression ``cosmology''.  This is discussed below, in Section \ref{totology}.  A considerable amount of the difficulties mentioned above stems, I argue, from this confusion. The discussion clarifies on the role of the quantum mechanical observer in cosmology, which is considered in Section \ref{god}.   In Section \ref{RQM}, I briefly describe the relational interpretation of quantum mechanics and some related aspects of quantum theory. In Section \ref{TIME}, I discuss the timeless aspect of the Wheeler-DeWitt equation.  In Section \ref{LQC}, I show how this perspective on quantum cosmology can be taken as an effective conceptual structure in the application of loop quantum gravity to cosmology.

\subsection{Cosmology is not about everything}\label{totology}

The expression ``cosmology'' is utilized to denote two very different notions.  The first is the subject of exploration of the cosmologists.  The second is the ``totality of things''. These are two very different meanings. 

To clarify why, consider a common physical pendulum. Its dynamics is described by the equation $\ddot q=-\omega q$, and we know well how to deal with the corresponding classical and quantum theory.  Question: does this equation describes ``everything'' about the physical pendulum?   The answer is obviously negative, because the pendulum has a complicated material structure, ultimately made by fast moving elections, quarks, and whatever, not to mention the innumerable bacteria most presumably living on its surface and their rich biochemistry...    The point is that the harmonic oscillator equation certainly does not describe the \emph{totality} of the physical events on the real pendulum: it describes the behaviour of one dynamical variable, neglecting everything else happening at smaller scales.  

In a very similar manner, cosmology  ---in the sense of ``what cosmologists actually study'' ---describes a number of large-scale degrees of freedom in the universe. This number may be relatively large: it may include all the measured CMB  modes, or the observed large-scale structures. But it remains immensely smaller than the total number of degrees of freedom of the real universe: the details of you reading now these words do not appear in any of the equations written by cosmologists. In strict sense, ``cosmology", defined as the object of study of the cosmologists, denote the \emph{large scale} degrees of freedom of the universe.  The fact that many shorter scale degrees of freedom are neglected is no different from what happens in any other science: a biologist studying a cat is not concerned with the forces binding the quarks in the nucleus of an atom in the cat's nose. 

There is however a different utilisation of the term ``cosmology" that one may find in some physics articles: sometimes it is used to denote an hypothetical science dealing with ``the totality of all degrees of freedom of Nature".    For the reason explained above, the two meanings of ``cosmology" are to be kept clearly distinct, and much of the conceptual confusion raised by quantum cosmology stem from confusing these two different meanings of the term.  For clarity, I use here two distinct terms: I call ``cosmology" the science of the large scale degrees of freedom of the Universe, namely the subject matter of the cosmologists. We can call ``totology" (from the latin ``totos", meaning ``all") the science -if it exists- of \emph{all} degrees of freedom existing in reality. 

The important point is that cosmology and totology are two different sciences. If we are interested in the quantum dynamics of the scale factor, or the emergence of the cosmological structures from quantum fluctuations of the vacuum, or in the quantum nature of the Big Bang, or the absence of a time variable in the Wheeler-DeWitt equation, we are dealing with specific issues in cosmology, not in totology.  

\subsection{The observer in cosmology}
\label{god}

The considerations above indicate that a notion of ``observer" is viable in cosmology.  In cosmology, the ``observer" is formed by ourselves, our telescopes, the measurement apparatus on our spacecrafts, and so on.  The ``system" is formed by the large scale degrees of freedom of the universe.  The two are clearly dynamically distinct.   The observer is not ``part of the system", in the dynamical sense.

Of course the observer is ``inside" the system in a spacial sense, because the scale factor describes the dimensions of a universe within which the observer is situated.  But this is no more disturbing than the fact that a scientist studying the large scale structure of the magnetic field of the Earth is situated within this same magnetic field.  Being in the same region of space does not imply being the same degrees of freedom. 

The notion of a dynamically external observer may be problematic in totology, but it is not so in conventional cosmology, and therefore there is no reason for it to be problematic in quantum cosmology.  

Actually, there is an aspect of the conventional presentation of quantum theory which becomes problematic: the idea that a system have an intrinsic physical ``state" which jumps abruptly during a measurement.   This aspect of quantum theory becomes implausible in cosmology, because the idea that when we look at the stars the entire universe could jump from one state to another is not very palatable.  But whether or not we interpret this jump as a physical event happening to the state  depends on the way we interpret the quantum theory.  This is why the application of quantum theory to cosmology bears on the issue of the interpretation of quantum mechanics.   There are some interpretations of quantum mechanics that become implausible when utilised for quantum cosmology.  But not all of them.   Below, I describe an interpretation which is particularly suitable for cosmology and which does not demand implausible assumptions such as the idea that our measurement could change the entire intrinsic state of the universe, 
as demanded by textbook Copenhagen interpretation.
As we shall see, in the relational interpretation of quantum mechanics the ``quantum state" is not interpreted as an intrinsic property of a system, but only as the information one system has about another.  There is nothing implausible if this changes abruptly in a measurement, even when the observed system is formed by the large scale degrees of freedom of the universe.


\section{Relational quantum mechanics}\label{RQM}

The relational interpretation of quantum mechanics was introduced in [\cite{Rovelli:1995fv}]
and has attracted the interest of philosophers such as Michel Bitbol, Bas van Frassen and  Mauro Dorato [see \cite{Dorato:2013ad} and references therein]. 

%

The point of departure of the relational interpretation is that the theory is about quantum events rather than about the wave function or the quantum state.   The distinction can be traced to the very beginning of the history of quantum theory: Heisenberg's key idea was to replace the notion of an electron continuously existing in space with a lighter ontology, the one given just  by  discrete tables of numbers.  The electron, in Heisenberg's vision, can be thought as ``jumping" from one interaction to another.
\footnote{Heisenberg gives a telling story about how he got the idea. He was walking in a park in Copenhagen at night. All was dark except for a few island of light under street lamps. He saw a man waking under one of those, then disappearing in the dark. Then appearing again under the next lamp. Of course, he thought, man is big and heavy and does not ``really" disappear:  we can reconstruct his path through the dark. But what about a small particle? Maybe what quantum theory is telling us is precisely that we cannot use the same intuitions for small particles. There is no classical path between their appearance here and their appearance there. Particles are objects that manifest themselves only when there is an interaction, and we are not allowed to fill up the gap in between. The ontology that Heisenberg proposes does not increase on the ontology of classical mechanics: it reduces it.  It is \emph{less},  not \emph{more}.  Heisenberg removes excess baggage from classical ontology and is left with a minimum necessary to describe the world.} 
 In contrast, one year later, Schr\"{o}dinger was able to reproduce the technical results of Heisenberg and his collaborators using a wave in space. Schr\"{o}dinger's wave evolved into the modern notion of quantum state.  What is quantum mechanics about: an evolving quantum state, as in Schr\"{o}dinger; or a discrete sequence of quantum events that materialise when systems interact?   

Relational quantum mechanics is an interpretation of quantum mechanics based on the second option, namely on Heisenberg's original intuition.  The advantage is that the quantum state is now interpreted as a mere theoretical booking device for the information about a system $S$ that a \emph{given} system $O$ might have gathered in the course of its past interactions with $S$.  Therefore the quantum state of $S$ is not intrinsic to $S$: it is the state of $S$ \emph{relative} to $O$. It describes, in a sense, the information that $O$ may have about $S$. No surprise if it jumps abruptly at a new interaction, because at each new interaction $O$ can gather new information about $S$. 

Thus, if we adopt this reading of quantum theory, there is no meaning in the ``the wave function of the entire universe", or ``the quantum state of everything", because these notion are extraneous to the relational  interpretation. A quantum state, or a wave function\footnote{The ``wave function" is the representation of the quantum state by means of a function on the spectrum of a complete commuting family of observables; for instance, a wave on configuration space.}, can only refer to the interactions between \emph{two} interacting subsystems of the universe. It has no more reality than the distributions of classical statistical mechanics: tools for computing.  What is real is not the quantum state: what is real are the individual quantum events.  For all these reasons, the relational interpretation of quantum mechanics is particularly suitable for quantum cosmology. 

Of course there is a price to pay for this remarkable simplification, as always the case with quantum theory. The price to pay here is that we are forced to recognise that the fact itself that a quantum event has happened, or not, must be interpreted as relative to a given system.  Quantum events cannot be considered absolute, their  existence is relative to the physical systems involved in an interaction.  Two interacting systems realise a quantum event relative to one another, but not necessarily relative to a third system.  This caveat, discussed in detail in Rovelli's original article and in the numerous philosopher's articles on the relational interpretation, is necessary to account for interference and to accommodate the equality of all physical systems. In fact, this is the metaphysical core of the relational interpretation, where a naive realism is traded for mature realism, able to account for Heisenberg's lighter ontology.

In cosmology, however, it is small price to pay: the theory itself guarantees that as long as the quantum effect in the interactions between quantum systems can be disregarded, different observing systems give the same description of an observed system and we are not concerned about this lighter ontology.   This is definitely the case in cosmology.\footnote{In spite of this lighter ontology, this interpretation of quantum theory takes fully the side of realism, in the sense that it assumes that the universe exists independently from any conscious observer actually observing it.  Consciousness, mind, humans, or animals, have no special role. Nor has any role any particular structure of the world (macroscopic systems, records, information gathering devices ...). Rather, the interpretation is democratic: the world is made by physical system, which are on equal ground and interact with one another. Facts are realized in interactions and the theory describes the probabilities of the outcome of future interaction, given past ones.  But this is a realism in weak sense of Heisenberg. In the moment of the interaction there is a real fact. But this fact exists relatively to the interacting systems, not in the absolute. In the same manner, two objects  have a well defined velocity with respect to one another, but we cannot say that a single object has an absolute velocity by itself, unless we implicitly refer to some other reference object. 
Importantly, the structure of quantum theory indicates that in this description of reality the manner in which we split the world in subsystems is largely arbitrary.  This freedom  is guaranteed by the tensorial structure of QM, which grants the arbitrariness of the positioning of the  boundary between systems which has been studied by Wigner: the magic of the quantum mechanics formalism is that we can split up the Hilbert space into pieces, and the formalism keeps its consistency.}

Under the relational reading reading of quantum theory, the best description of reality we can give is the way things affect one another. Things are manifest in interactions (quantum relationalism). Quantum theory describes reality in terms of facts appearing in interactions, and relative to the systems that are interacting.  Cosmology describes the dynamics ---possibly including the quantum dynamics--- of the large-scale degrees of freedom of the universe and the way these are observed and measured by our instruments.  

Before closing this Section on the relational interpretation, I discuss below two important notions on which this interpretation is based: \emph{discretness} and quantum \emph{information}. I give also, below, a discussion on the role of the \emph{wave function} quantum mechanics, and on its common overestimation.

%
%

\subsection{Discreteness}\label{Discreteness}

Quantum mechanics is largely a discovery of a very peculiar form of discreteness in Nature.  Its very name refer to the existence of peculiar discrete units: the ``quanta".  Many current interpretations of quantum mechanics underemphasise this discreteness at the core of quantum physics, and many current discussion on quantum theory neglect it entirely. 
Historically, discreteness played a pivotal role in the discovery of the theory:
\begin{itemize}
\item{1900 Planck:}                                           finite size packets of energy $E=h\nu$
\item{1905 Einstein:}                                         discrete particles of light
\item{1912 Bohr:}                                              discrete energy levels in atomic orbits
\item{1925 Heisenberg:}               tables of numbers
\item{1926 Schr\"odinger:} discrete stationary waves
\item{1930 Dirac:}                                 state space and operators with possibly discrete spectra
\item{Spin:} discrete values of angular momentum
\item{QFT:} particles as discrete quanta of a field
\end{itemize}

As all these examples indicate, the scale of all these examples of discreteness is always set by $\hbar$. But what is actually discrete, in general, in quantum mechanics?
It is easy to answer by noticing that $\hbar$ has the dimensions of an action and that the (Liouville) volume of the phase space of any system has always the dimension of an action (per degree of freedom). The constant $\hbar$ sets a unit of volume in the phase space of any system.  The physical discovery at the basis of quantum theory is the impossibility of pinpointing the classical state of a system with a precision superior to such a unit volume.  

If we have made some measurements on a system, with a given accuracy, and, consequently, we know that the system is in a certain region $R$ of its phase space, then the quantum theory tells us that there are only a \emph{finite} number of possible (orthogonal, namely distinguishable) states in which the system can be, given by 
\begin{equation}
N=\frac{Volume(R)}\hbar~.
\end{equation}
 \begin{wrapfigure}{r}{5,5cm}
\begin{center}
\vspace{-2em}
\includegraphics[height=35mm]{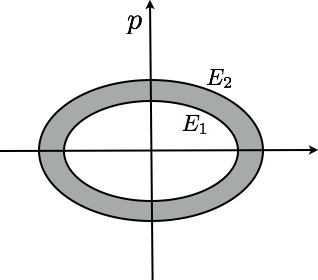}
\end{center}
\caption{In the shaded region $R$ there is an infinite number of classical states, but only a finite number of quantum ones.}
\label{uno}
\end{wrapfigure}
Let's consider a simple example: say that measuring the energy  $E$ of an harmonic oscillator we learn that $E \in [E_1,E_2]$. This determines a region $R$ of phase space, with volume $V=\int_{E \in [E_1,E_2]}dp\,dq=2\pi\frac{E_2-E_1}{\omega}$ (see Fig.\ref{uno}). There is  an infinite number of possible classical states, and an infinite number of possible values that the energy can actually have. In the quantum theory -that is, in Nature- the energy cannot take all these values, but only a finite number, which is $I=\frac{E_2-E_1}{\hbar\omega}=\frac{V}\hbar$. This is the maximum number of orthogonal states compatible with the previous measurement. 
\newpage
The example shows that a region of phase space is the specification of a certain amount of information we have about a system.  This is general: quantum mechanics teaches us that in every finite region of phase space we can only accommodate a finite number of orthogonal states, namely that there is a finite information that we can extract from any finite region of phase space.

The same is true in quantum field theory. The quanta of a field are discrete particles (Dirac) and this is precisely a manifestation of this same discreteness; in particular, the finiteness of the spectrum of the energy of each mode. \\
\indent
Discreteness, in this sense, is the defining property of all quantum system. The discreteness scale is set by $\hbar$, an action, or phase-space volume. 

\subsection{Information}

The notion of information useful in quantum theory is Shannon's \emph{relative information}, which is defined as follows. Given two systems, such that we can find the first system in $N_a$ states and the second in $N_b$, we say that the two have information about one another if, because of physical constraint, we can find the combined state in a number $N_{ab}$ of states which is smaller than $N_a\times N_b$. The relative information is then defined by 
\begin{equation}
         I = \log_2(N_a\times N_b)- \log_2(N_{ab}) 
\end{equation}
The utility of this definition of information is that there is nothing mental or subjective about it: it is simply a measure that physics establishes between two degrees of freedom.  For instance, as long as my pencil is not broken, each extreme of the pencil has information about the other, because the two can be in a smaller number of places than two separated objects.   Knowledge of the position of one give some information about the position of the other. 

The existence of this correlation is a measurable property of the combined system. If we have information about a system, we can make predictions about the outcome of future interactions with it. We can call ``relevant" information that portion of the information that we have about a system which is not redundant in view of the such predictions.  In the relational interpretation, physics is the theory of the relevant relative information that systems can have about one another. In  [\cite{Rovelli:1995fv}], two basic postulates were proposed, meant to capture the physical content of quantum theory: 
\begin{description}
\item[Postulate I] There is a maximum amount of relevant information that can be extracted from a system.
\item[Postulate II] It is always possible to acquire new relevant information about a system~.
\end{description}
Remarkably, the entirety of the quantum mechanical formalism (Hilbert spaces, self adjoint operators, eigenvalues and eigenstates, projection postulate...) can be recovered on the basis of these two postulates, plus a few other more technical assumptions.  The effort of understanding the physical meaning of these additional postulates, and to provide a mathematically rigorous reconstruction theorem has been developed by a number of authors and is still in course [\cite{Grinbaum:2003kb,Hoehn:2014uua}].

A given system $S$ can be characterised by a set of variables. These take values in a set (the spectrum of the operators representing them, possibly discrete). In the course of an interaction with a second system $O$, the effect of the interaction on $O$ depends on the value of one of these variables.  We can express this by saying that the system $O$ has then the information that this variable of the system $S$ has the given value.  This is one of the elementary quantum events that quantum theory is concerned with.  The first postulate captures an essential aspect of quantum theory: the impossibility of associating a single point in phase space to the state of a system: to do so, we would need an infinite amount of information.  The maximum localisation in phase space is limited by $\hbar$ which determines the minimal physical cell in phase space.  The information about the state of a system is therefore limited.  The second postulate distinguishes quantum systems from classically discrete systems: even if finite, information is always incomplete, in the sense that new information can be gathered, at the price of making previous information irrelevant for predicting the future interactions: by measuring the $L_x$ component of angular momentum we destroy information we previously had on the $L_z$ component. 

\subsection{The role of the wave function}

If the real entities in quantum theory are discrete quantum events, what is the quantum state, or the wave function, which we commonly associate to an evolving system on many applications of quantum theory?  

A good way to make sense of a quantity in a theory is to relate it to the corresponding quantity in an approximate theory, which we understand better for historical reasons. We can therefore investigate the meaning of the quantum mechanical wave function by studying what it becomes in the semiclassical limit. As well know, if $\psi(x,t)$ is the wave function of a Newtonian particle:            
                                          \be
                                          \psi(x,t)\sim e^{\frac i\hbar S(x,t)}
                                          \ee
                                              where   $S (x,t)$ is the Hamilton function, the Schr\"odinger equation becomes in the classical limit:
                                          \be
                                          -i\hbar\frac\partial{\partial t}\psi(x,t)
                                          = - \frac{\hbar^2}{2m} \frac{\partial^2}{\partial x^2}\psi(x,t)+V(x)\psi(x,t)
                                          \ee     
                                          
                                           \be \!\! \!\! \!\! \!\! \!\! \!\!\longrightarrow \ ~
                                          \frac\partial{\partial t}S(x,t)
                                          = \frac{1}{2m} \frac{\partial^2}{\partial x^2}S(x,t)+V(x)
                                          \ee   
                                                                                which is the Hamilton-Jacobi equations.                         

The Hamilton function is a theoretical device, not something existing for real in space and time. What exists for real in space and time is the particle, not its Hamilton function. I am not aware of anybody suggesting to endow the Hamilton function with a realistic interpretation. 

So, why therefore would  anybody think of doing so with its corresponding quantum object, the wave function?  A interpretation of quantum theory that considers the wave function a real object sounds absurd: it is assigning a reality to a calculation tool.  If I say: ``Tomorrow we can go hiking, or we can go swimming", I am not making a statement implying that tomorrow the two things will both happen: I am only expressing by ignorance, my uncertainty, my lack of knowledge, my lack of information, about tomorrow.  Perhaps tomorrow neither program would realise, but not both. The wave function of a particle expresses probabilities, and these are related to our lack of knowledge. It does not mean that the wave function becomes a real entity spread in space, or, worse, in configuration space. 
The wave function (the state) is like the Hamilton-function: a computational device, not a real object [\cite{Durr:1995du}].

In science, abstract concepts have sometimes been recognized for ÒrealÓ entities.  But  often the opposite has happened: a misleading attitude has come from realism barking at the wrong tree: examples are the crystal spheres, the caloric, the ether... A realist should not be realist about the quantum state, if she wants to avoid the unpalatable alternative between the Scilla of the quantum collapse or the Cariddi of the branching many worlds.   We can make sense of quantum theory with a softer realism, rather than with an inflated one.  The wave function is abut the relative information that systems have about each others: it is about information or lack of information, it is about what we can expect about the next real quantum event.

\subsection{Relevance for quantum cosmology}

As argued in the previous section cosmology, in the conventional sense, is the study of the dynamics of these large scale degrees of freedom. It is not about everything, it is about a relatively small number of degrees of freedom.   It is essentially  based on an expansion in modes and neglect ÒshortÓ wavelength modes, where  ``short'' includes modes of millions of light years. It describes the way the large nodes affect all the (classical) rest, including us.  The dynamics of these degrees of freedom is mostly classical, but it can be influenced by quantum mechanics, which introduces discreteness and lack of determinisms, in some regimes such as the early universe.  This quantum aspects of the dynamics of the universe at large can be taken into account by standard quantum mechanics.   The ``observer" in this case, is simply formed by ourselves and our instruments, which are not part of the large scale dynamics.  The quantum state of the system represents the information we have gathered so far about the large scale degrees of freedom.  As usual in quantum theory, we can often ``guess" aspects of these states, completing our largely incomplete observations about them.  

As far as the other interpretation of the term ``cosmology" is concerned, namely as totology, at the light of the relational interpretation of quantum theory, it is perhaps questionable whether a coherent ``quantum totology" make any sense at all.  If we define a system as the totality of anything existent, it is not clear what is the meaning of studying how this system appears to an observer, since by definition there is no observer. The problem is open, but it is not much related to the concerns of the real-life cosmologists, or to questions such as what happened at the big bang. 

We can then use standard quantum theory, for instance in the form of a Hilbert space for those large degrees of freedom, observables that describe the physical interaction between the large degrees of freedom and our telescopes. For instance we can measure the temperature of the CMB.  This can be done with the usual conceptual tools of quantum theory. The ÒsystemÓ does not include the observer. The ÒobserverÓ in cosmology is indeed the majority of stuff in the universe: all the degrees of freedom of the universe except for the large scale one. This definition of  cosmology eliminates any problem about the observer.  A measurement does not need to ``affect the large scale universe". It only affects our information about it. 


\section{Quantum gravity and quantum cosmology}\label{LQC}

As best as we know, gravity is described by general relativity. The peculiar symmetries of general relativity add specific conceptual issues to the formulation of quantum cosmology.  Among these, is the fact that instead of a Schr\"odinger equation, evolving in a time variable $t$, we have a Wheeler-DeWitt  equation, without any explicit time variable. 

It is important to distinguish different issues.  The discussion of the two previous Sections would not change if gravity was described by Newton's theory and our main quantum dynamical equation was a Schr\"odinger equation.  The distinction between cosmology and totology would not change. Contrary to what often stated, the lack of a time variable in the Wheeler-DeWitt equation has no direct relation with the existence or the absence of an external quantum observer, because the problem giving a quantum description of the totality of things would be the same also if gravity was Newtonian.  

The additional, specific, problem raised by general relativity is how to describe \emph{evolution} in the presence of general covariance.  The solution, on the other hand, is well known and is very similar to the solution of the problem of the observer discussed above: quantum theory describes the state of a system \emph{relative} to another --arbitrary-- system.  General relativity describes the evolution of some observables \emph{relative} to other --arbitrary-- observables.  

This solution is recalled below. 

\subsection{The relational nature of general relativity}

The celebrated idea of Einstein in 1915 is that spacetime {\it is} the gravitational field. This imply that
spacetime is a dynamical field. General relativity does not imply that the gravitational field is particularly  different from other fields. It implies that all physical fields do not live in spacetime: rather, the universe is made by several fields, interacting with one another. One of these is the gravitational fields.  The description of some specific aspect of a configuration of this field is what we call geometry. 

In the region of the universe where we live, a good approximation is obtained by neglecting both the dynamics of the gravitational field and its local curvature, and use the gravitational fields as a fixed background with respect to which we can define acceleration and write Newton's second law. 

Localisation is relative, to other dynamical objects, including the gravitational field.  This is also true for temporal localisation.  In general relativity there is no fixed background structure, nor a preferred time variable with respect to which events are localised in time. In Newtonian physics there is a time along which things happen; in general relativity there is no preferred time variable. Physical variables evolve with respect to one another.  For instance, if we keep a clock at a fixed altitude and we throw a second clock upward so that it raises and then falls back next to the first clock, the readings of the two clock will then differ. Given sufficient data, general relativity allows us to compute the value of the first $t_1$ as a function of the second $t_2$, or vice versa. None of the two variables $t_1$ or $t_2$ is a more legitimate ``time" than the other. 

This observation can be formalised in terms of the notion of \emph{partial observable} [\cite{Rovelli:2013bf}], which provides a clean way to deal with the peculiar gauge structure of general relativity.  

In the example above, the two variables $t_1$ and $t_2$, representing the reading of the two clocks, are partial observables.   Both quantities can be measured, but none of the two can be predicted, of course, because we do not know ``when" an observation is made.   But the value of each can be predicted once the other is known.  The theory predicts (with sufficient information) the  \emph{relation} between them.  

In the Hamiltonian formulation of the theory, the dynamics of general relativity is generated by constraints, as direct consequence of the absence of background. The solution of those constraints code the evolution of the system, without external time with respect to whom evolution can be described [\cite{Rovelli:1989jn}].  Partial observables are functions on the extended phase space where constraints are defined. On the constraint surface, the constraints generate orbits that determine relations between partial observables.  These relations express the classical dynamical content of the theory.   

The prototypical example of a partial observable which can be measured but not predicted is the conventional time variable of Newtonian physics: a quantity that we routinely determine (looking at a clock) but we can not predict from the dynamics of the system. 

The physical phase space is the space of these orbits. A point of phase space cannot be interpreted as the characterisation of the state of the system at a given time, because the theory has no notion of time.   Rather, it can be interpreted as a way for designing a full solution of the equations of motion.  But the dynamical information is not just in the physical phase space: it is in the relation between partial observables that each orbits determines. 

In the quantum theory, the strict functional dependence between partial observables determined by the classical dynamics is replaced by transition amplitudes and transition probabilities.  Thus, in a general covariant theory, physical observations are given by the transition amplitudes between eigenstates of partial observables.   Formally, these are given by the matrix elements of the projector on the solution of the Wheeler-DeWitt operator between these states, which are defined on the same extended Hilbert space on which the Wheeler-DeWitt operator is defined.

\subsection{Time in quantum cosmology}\label{TIME}

In cosmology it is often convenient to choose one of the variables and treat it as a ``clock variable", that is, an independent variable with respect to which study the evolution of the others.  The choice is dictated by convenience, and has no fundamental significance whatsoever.  For instance, there is no particular reason for choosing an independent variable that evolves monotonically along the orbits, or that defines a unitary evolution in the quantum theory. 

In the case of a homogeneous and isotropic cosmology, where only one degrees of freedom is considered, at least a second one is needed, to be used as a clock. A common choice is  a massless scalar field. More ingenious strategy can be implemented with enough degrees of freedom: for instance in the Bianchi I cosmology where the three spacial direction can evolve independently, one spacial direction can be taken to play the role of time.  One should be careful to deal with regions where  the chosen time fails to be monotonic. For instance the scale factor, that is implicitly taken as clock in many models of quantum cosmology, is not monotonic if there is a recollapse.

Consider the study of the quantum mechanics of the scale factor $a$ plus a single other degree of freedom, say a scalar field $\phi$ representing the average matter energy density. The two variables $\phi$ and $a$ are partial observables. Predictions are extracted from their relative evolution of   realizing Einstein's relationalism [\cite{Ashtekar:2007tv}].

In loop quantum cosmology, in particular, the dynamics studied includes effects  of the fundamental \emph{spacetime discreteness} revealed by loop quantum gravity, using the technique of loop quantization. Among the results of the theory are the generic resolution of curvature singularities and the indication of the existence of a bounce replacing the initial singularity: a classical contracting solution of the Einstein's equations can be connected to an expanding one via a quantum tunneling. The bounce is a consequence of the Heisenberg relations for gravity, in the same way in which for an atomic nucleus those prevents the electrons to fall in [\cite{Bojowald:2001xe,Ashtekar:2006rx}]. In the easiest case of a FLRW universe with no curvature, the effective equations provide a simple modification of the Friedmann equation that is:
 \be
\left( \frac{\dot a}a \right)^2=\frac{8\pi G}3 \rho\left(1-\frac\rho{\rho_c}\right), 
\label{cfe}
\ee
where $\rho_c$ is the critical density at which the bounce is expected and that can be computed to correspond roughly speaking to the Planck density. The effects of the bounce on standard cosmological observables, such as CMB fluctuations, have been lengthily studied, see for instance [\cite{Ashtekar:2015bs}].

There results are of course tentative and wait for an empirical confirmation, but the standard difficulty regarding time evolution does not plague them.


\subsection{Covariant loop quantum gravity}

If we move to the quantum description of a small number of degrees of freedom, as in the last Section, to a full quantum theory of gravity, which is ultimately needed in quantum cosmology, some interesting structures appear. 

The main point is that in the absence of time we have to modify the notion of ``physical system" used in Section \ref{RQM}. The reason is that the notion of quantum system implies a permanence in time which looses meaning in the fully covariant theory: how do we identify the ``same system" at different times in a covariant field theory? 

The solution is to restrict to local processes.  The amplitudes of quantum gravity can be associated to finite spacetime regions, and the states of quantum gravity to the boundary of these regions.  In fact, in quantum gravity we may even identify the notion of a spacetime region with the notion of {\em process}, for which we can compute transition amplitudes.  The associated transition amplitudes depend on the eigenstates of partial observables that we can identify with spacial regions bounding the spacetime region of the process.  In particular, loop quantum gravity gives a mathematical precise definition of the state of space, the boundary observables, and the amplitude of the process, in this framework. 
The possibility of this ``boundary" formalism [\cite{Oeckl:2003vu}] stems from a surprising convergence between general relativity and quantum theory, which we have implicitly pointed out above. 

We can call ``EinsteinÕs relationalism'' the fact that in general relativity localization of an event is relative to other events.  
We can call ``quantum relationalism" the fact that quantum theory is about the manner a system affects another system. In Bohr's quantum theory, the attention was always between the quantum system and the classical world, but we have seen that relational quantum theory allows us to democratise this split and describe the influence of any system on any other.  These two relationalisms, however, appear to talks one another, because of the locality of all interactions [\cite{Vidotto:2013jia}]. 

Indeed, one of the main discovery in modern physics is locality: interactions at distance of the Newton's kind don't seem to be part of our world. In the standard model things interact only when they ``touch'': all interactions are local. But this  means that objects in interactions should be in the same place: interaction require localization and localization requires interaction.  To be in interaction correspond to be adjacent in spacetime and vice versa: the two reduce to one another.

\hspace{-2cm}
{\small
\begin{table*}[htdp]
\begin{tabular}{ccc}
Quantum relationalism          & $\longleftrightarrow$ &         Einstein's relationalism \\
\hspace{-6mm}  Systems interact     with other systems               & $\longleftrightarrow$&                Systems are located    wrt other systems\\
            Interaction = Localization     &$\longleftrightarrow$ &   Localization =  Interaction
\end{tabular}
\label{default}
\end{table*}%
}

Bringing the two perspectives together, we get to the boundary formulation of quantum gravity: the theory describes processes and their interactions. The manner a process affects another is described by the Hilbert state associated to its boundary. The probabilities of one or another outcome are given by the transition amplitudes associated to the bulk, and obtained from the matrix elements of the projector on the solutions of the Wheeler De Witt equation.

Let us make this more concrete. Consider a process such as the scattering of some particles at CERN. If we want to take into account the gravitational filed, we need to include it as part of the system. In doing quantum gravity, the gravitational field (or spacetime) is part of the system.
Distance and time measurements are field measurements like the others in general relativity:  they are part of the boundary data of the problem.

Thinking in terms of functional integrals, we have to sum over all possible histories, but also all possible geometries associated to a given finite spacetime region.

In the computation of a transition amplitude, we need to give the boundary data of the process that are for instance the position of a particle at an initial and a final time. We use rods and clocks to define them. But those measure geometrical informations that are just value of the gravitational field. Everything we have to give is the value of the fields on the boundary. This includes the gravitational fields from which we can say how much time have passed and the distance between the initial and the final point. Geometrical and temporal data are encoded in the boundary state, because this is also the state of the gravitational field, which is the state of spacetime. 

 \begin{figure}[h]
\begin{center}
\includegraphics[height=30mm]{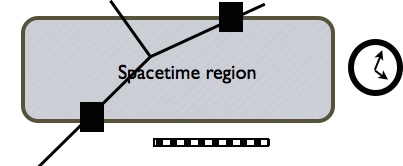}
\end{center}
\caption{Boundary values of the gravitational field  
=    geometry of box surface  =    distance and time separation of measurements
}\end{figure}

This clarifies that in quantum gravity a process is a spacetime region.

Now, in we have seen that in relational quantum mechanics we need systems in interaction. What defines the system and when is it interacting? For spacetime, a process is simply a region of spacetime. Spacetime is a quantum mechanical process once we do quantum gravity. Vice versa, this now helps us to understand how to do quantum gravity.  

Notice that from this perspective quantum gravitational processes are defined locally, without any need to invoke asymptotic regions. Summarising: \\

{\small \begin{center}
{\bf Quantum dynamics of spacetime}
\begin{tabular}{ccc}
Processes    &$\to$&   Spacetime regions\\
States    &$\to$&  Boundaries   =  spacial regions \\
Probability    &$\to$&   Transition amplitudes\\
Discreteness   &$\to$&  Quanta of space\\[1em]
\end{tabular}
\end{center}
\label{default}}
\subsection{Discreteness in quantum gravity}

In Section \ref{Discreteness}, I have discussed how $\hbar$ give us a unit of action in phase space, and a conversion factor between action and information. In gravity the phase space is the one of possible 4-dimensional geometries, and there is the Newton constant $G$ which transform regions of phase space in lengths. What kind of discreteness does this imply? 

The answer is the well-known Planck length, originally pointed out by Bronstein while debating a famous argument on field's measurability by Landau, in the case of the gravitational field [\cite{Rovelli:2014ssa}].

The argument is simple: in order to check what happens in a small region of spacetime, we need a test particle. The smaller the region the more energetic the particle should be. Until energy  curves spacetime to form a black hole, whose horizon beyond which nothing can be seen is larger than the original region we wanted to prove. Because of this, it is not possible to probe scales smaller than  the Planck length $\ell_{P\ell}$. This is the core of core quantum gravity: the discovery that there is a minimal length.
 
These handwaving semiclassical arguments can be made rigorous in the loop theory studying the phase space of general relativity and the corresponding operators. 
Geometrical quantities, such as area, volumes and angles, are function of the gravitational field that is promoted to operators. Their  discrete spectra describes a spacetime that is granular in the same sense in which the electric field is made of photons. For instance, the spectrum of the area  can be computed and it results to be discrete 
[\cite{Rovelli:1994ge}]:
\be
A=8\pi \ell_{P\ell} \sqrt{j(j+1)}, ~ j\in\frac{\mathds N}2
\ee
where $j$ is a half integer, similarly to what happens for the angular momentum. This has a minimal eigenvalue: a minimal value for the area. 

Loop quantum gravity describes how these quanta of spacetime interacts one another. The notion of geometry emerges only from  the semiclassical picture of these interactions. 
Formally, the theory is defined as follows. Every quantum field theory can be given in terms of a triple $(\cal H, \cal A, W)$: 
respectively a Hilbert space where the states live, an algebra of operator, and the dynamics defined in the covariant theory by a transition amplitude. 

The interactions with the field manifests the discreteness of quantum mechanics: the fundamental discreteness appears in the presence of particles, that are just the quanta of a field, and in the spectrum of the energy of each mode of the field. The same structure applies to loop quantum gravity: states, operators and transition amplitudes can be properly defined [\cite{Rovelli:2014ssa}] and there is a fundamental discreteness: the granularity of spacetime,  yielded by the discreteness of the spectrum of geometrical operators. The geometry is quantized: eigenvalues are discrete and operators do not commute.
Nodes carry discrete quanta of volume ({\em quanta of space}) and the links  discrete quanta of area.
Area and volume form a complete set of commuting observables and have discrete spectra. 

States in loop quantum gravity are associated to graphs characterized by $N$ nodes and $L$ links. They can be thought as analogous to the $N$-particle states of standard quantum field theory, but with some further extra-information given by the links, that  turn out to be adjacency relations coding which ``quanta of spacetime'' are interacting one another. These quanta {\it are} spacetime, they do not live in spacetime. The graphs are colored with quantum numbers, i.e. spins, forming the mathematical object called ``spinnerwork'' 
[\cite{Penrose:1971sn}].
Penrose's ``spin-geometry'' theorem connects the graph Hilbert space with the description of the geometry of a cellular decomposition of spacetime. 
\\

\begin{minipage}[b]{0.4\textwidth}
\includegraphics[width=\textwidth]{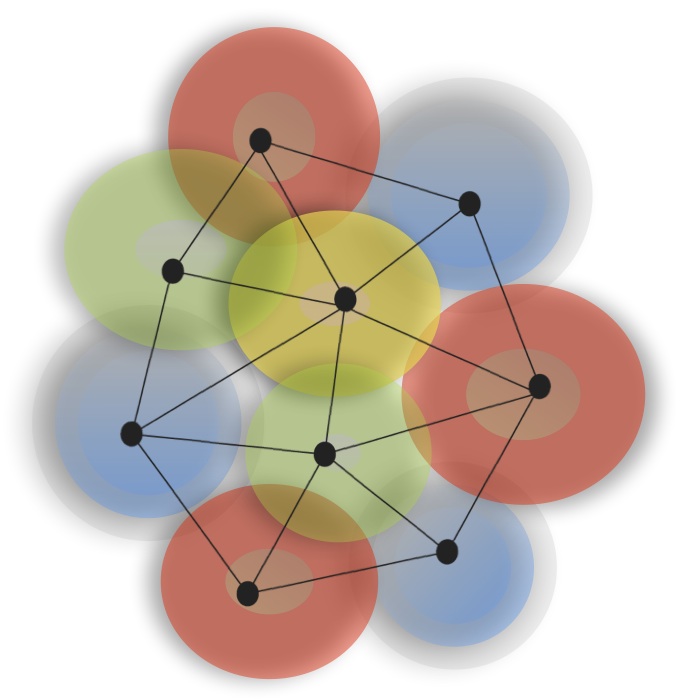} \\  \centering Spinnetwork\end{minipage}
\hskip15mm
\begin{minipage}[b]{0.3\textwidth}
 \includegraphics[width=\textwidth]{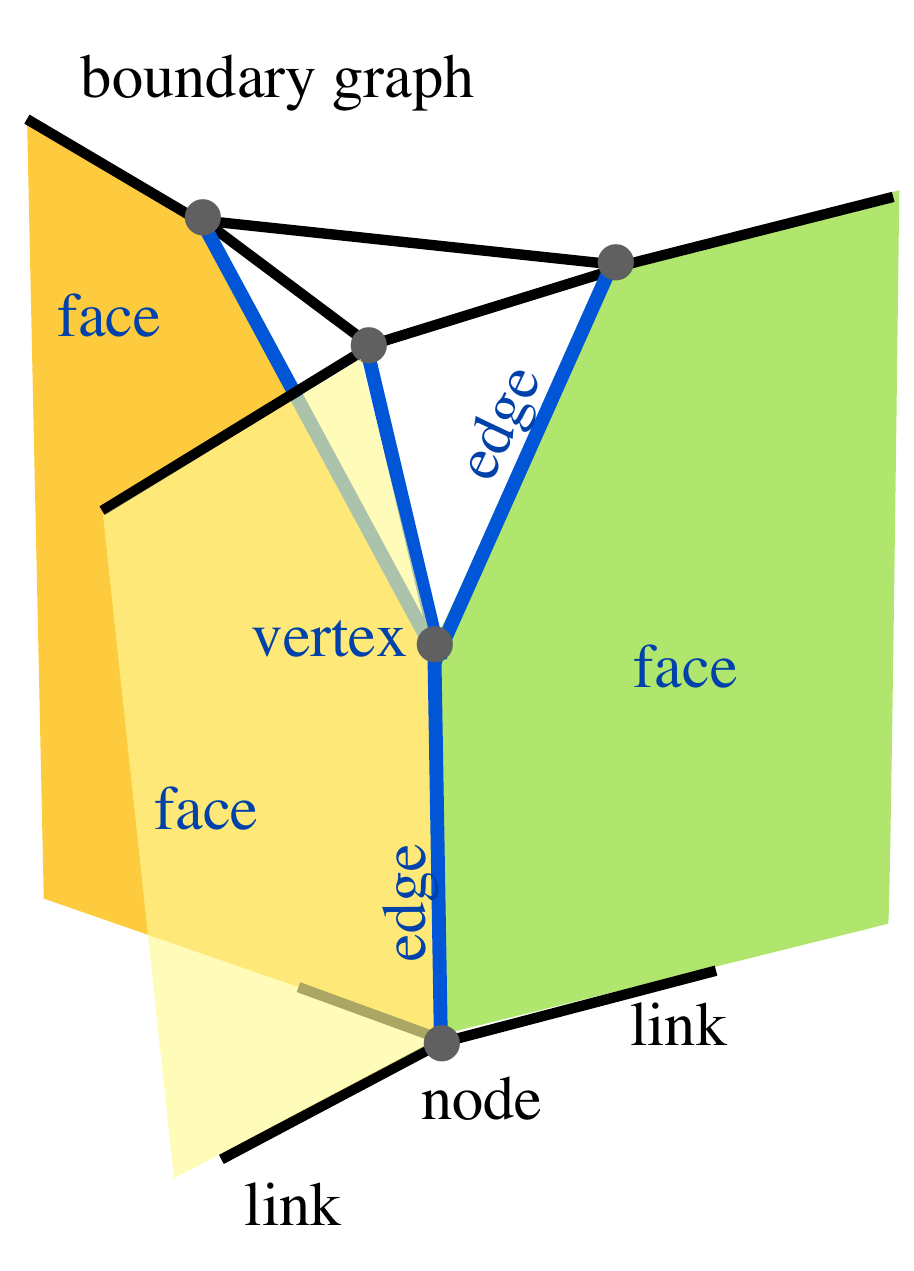} \\  \centering Spinfoam
  \end{minipage}
\\

Notice that the full Hilbert space of the theory is formally defined in the limit of an infinite graph, but the physical theory is capture by a finite graph in the same way in which the Fock space is truncated to $N$ particles. The truncation to a given finite graph captures the relevant degrees of freedom of the state we are interested to describe, disregarding those that need a ``larger'' graph to be defined.

A transition amplitude that represent a history of the geometry, in terms of graph states become a history of the boundary graphs, or a ``spinfoam''. In a spinfoam quanta/nodes and links/relations get transformed into new configuration by the action of interaction vertex in the bulk. A link span a face trough its history. This is the way of picturing a history of the quanta when these quanta makes up spacetime themselves. This yields a ontological unification where all that exists are \emph{covariant quantum fields} [\cite{Vidotto:2014cg}].

\subsection{Spinfoam Cosmology}

On a compact space we can expand the dynamical fields in discrete modes. The truncating of the theory to a finite number of modes defines an approximation to the full theory.   This is neither a large scale approximation nor a a short scale approximation, because the total space can still be very large or very small, as its scale is determined by the lowest modes of the gravitational field.  Rather, the approximation is in the ratio between the largest and the smallest relevant wavelengths considered.  

The graph expansion of the spin-network formulation of loop quantum gravity can be put in correspondence with this mode expansion of the fields on a compact space [\cite{Rovelli:2008ys}]. A truncation on a fixed graph corresponds then to a truncation in the mode expansion.  The truncation  provides a natural cut off of the infinite degrees of freedom of general relativity down to a finite number.  Choosing a graph, we disregard the higher modes of this expansion. The truncation defines an approximation viable for gravitational phenomena where the ratio between the longest and the shortest wavelength is bounded. 

Since this is neither an ultraviolet nor an infrared truncation, what is lost are not wavelengths shorter than a given length, but rather wavelengths $k$ times shorter than the full size of physical space, for some integer $k$.  

This approximation is useful in cosmology.   According to the cosmological principle, the dynamics of a homogeneous and isotropic space provides a good first order approximation to the dynamics of the real universe. Inhomogeneities can be disregarded in a first approximation.  Notice that the approximation is not just a large scale approximation, because the universe may be small at some point of its evolution. Rather, the truncation is in the ratio between the scale of the inhomogeneities and the scale factor. At lowest order, we consider the dynamics of the whole universe as described solely by the scale factor, this ratio is unit and a single degree of freedom is sufficient. We can then recover the rest of the theory adding degrees of freedom progressively. In the context of spin-foam cosmology, this can be obtained refining progressively the graph. 

A graph with a single degree of freedom is just a single node: in a certain sense, this is the case of usual Loop Quantum Cosmology. To add degrees of freedom, we add nodes and links with a coloring [\cite{Borja:2011ha}]. These further degrees of freedom are a natural way to describe inhomogeneities and anisotropies. 

Therefore a single graph provides a useful calculation tool in cosmology. It is possible to generalize the spinfoam technics for cosmology to large graphs.  In a \emph{regular} graph, which corresponds to a regular cellular decomposition, node and links become indistinguishable, and we obtain back the unique FLRW degrees of freedom [\cite{Vidotto:2011qa}].  For an arbitrarily large regular graph, we can define coherent states and peace them on an homogeneous and isotropic geometry, to represent macroscopic cosmological states.

Once we interpret the graph states as describing a cosmological evolution, we can compute cosmological transition amplitudes [\cite{Vidotto:2010kw,Bianchi:2010zs,Bianchi:2011ym}].
This transition amplitude makes concrete the notion of a sum over possible histories, namely all the possible 4-geometries compatible with the given 3-dimensional states on the boundary.\footnote{In the lowest approximation, the classical theory expresses the dynamics as a relation between the scale factor and its momentum. Consequently, in the quantum theory, at the first order in the vertex expansion the probability of measuring a certain ``out'' coherent state does not depend on the ``in'' coherent state. In other words, at the first order the probability is dominated by the product of the probabilities of each state to exist. Each term is given by a sum over all the possible 4-dimensional geometry compatible with the state representing a given 3-dimensional geometry. 
This is exactly the ``spinfoam version'' of the 
wave function of the universe [\cite{Hartle:1983ai}].}

The advantage of this formalism is that it is fully Lorentzian, the amplitudes are infrared and ultraviolet fine, and they have a good classical behavior as they result to be peaked on solutions of classical general relativity. Since the theory is non-perturbative, we are allowed to use these equations in the deep quantum regime, where a perturbative calculation would exit its domain of validity. The hope is to  obtain a full description of the quantum fluctuations at the bounce that replace the classical singularities.

Once again, the theory is tentative and may have technical difficulties, but there are no conceptual obstacles, if we adopt a fully relational perspective.


\section{Conclusions}

The world can be described in terms of facts. Facts happen at interactions, namely when a system affects another system, or, in a covariant theory, when a process affects another process.  Relational quantum mechanics is the understanding of quantum theory in these terms.   The resulting ontology is relational, characterizes quantum theory and is more subtle than that  of classical mechanics.   

Attributing ontological weight to the wave function is misleading. A quantum states are only the coding  of past events happened between two systems, or, in a generally covariant theory, between two processes.  The way a system will affect another system in the future is probabilistically determined by the manner it has done so in the past, and physics is about the determination of such probabilistic relations. The quantum state codes the information relevant for this determination. 

The amount of information is discrete in quantum theory.  The minimal amount of information is determined by the Planck constant.  The core of quantum theory is the discreteness of the information. 

Cosmology is not about everything. It is about a few large scale degrees of freedom.  It is based on an expansion in modes and neglects ``short" wavelength modes, where  ``short" means millions of light years.
Accordingly, in quantum cosmology the ÒsystemÓ does not include the observer. The observer is ourselves and our instruments, which are at a scale smaller than the cosmological scale. 

In contrast,  it is not clear whether a quantum theory of everything --a ``totology"--  makes sense, because quantum theory describes how a system affects another system. This question, however, has no bearing on standard quantum cosmology, which describes how the large scale degrees of freedom affect our instruments.  

Similarly, a preferred notion of time is not needed in quantum cosmology because the theory is about relations between partial observables, and dynamics is the study of these correlations.  

Quantum gravity is the theory of the existence of a minimal length. Spacetime is (quantum) discrete.
Since spacetime is dynamical, processes are spacetime regions. Neither space nor time are defined Òinside a processÓ. Thanks to this, the application of quantum gravity to quantum cosmology, cures the initial singularity and may lead to observable effects.  

Using this relational understanding of quantum mechanics and of evolution, a coherent and consistent formulation of quantum cosmology is possible.   

\vskip3cm

\begin{acknowledgements}
I acknowledge support of the Netherlands Organisation for ScientiÞc Research (NWO) under the Veni program.
\end{acknowledgements}

\end{document}